\documentclass[multphys,vecphys]{svmult}

\usepackage{makeidx}         
\usepackage{graphicx}        
\usepackage{multicol}        
\usepackage[bottom]{footmisc}

\makeindex             

\begin{document}

\title*{Helium and Iron in X-ray galaxy clusters}
\author{Stefano Ettori}
\institute{INAF, Osservatorio astronomico, via Ranzani 1, 40127 Bologna (Italy)
\texttt{stefano.ettori@oabo.inaf.it} }
%
%
\maketitle

\section{The metals in X-ray galaxy clusters}
\label{sec:1}

\begin{figure}[hb]
\centering
\includegraphics[height=5cm]{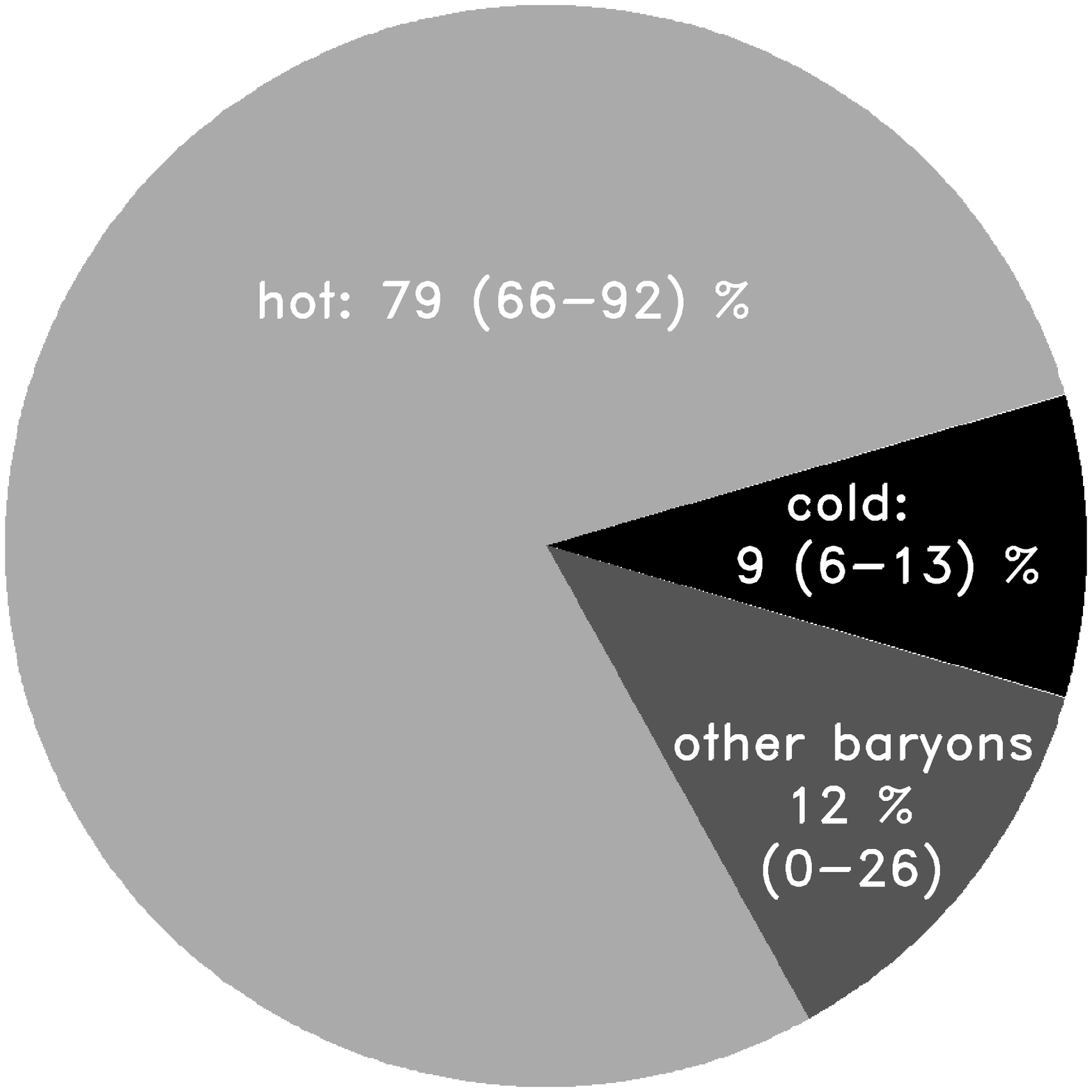}
\includegraphics[height=5cm]{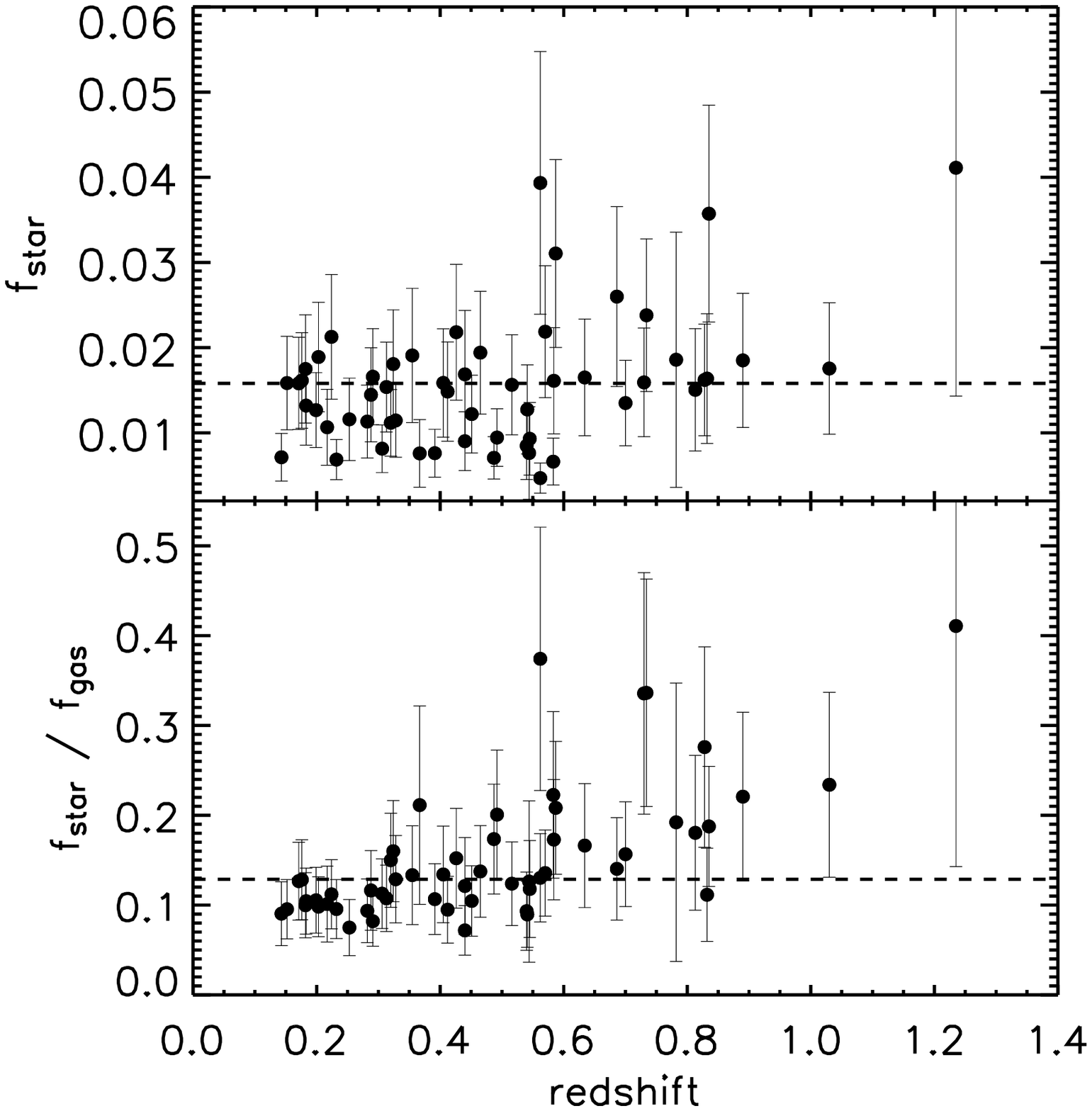}
\caption{(Left) Updated version of the cluster baryonic pie
presented in \cite{e03} and obtained by using 
$f_{\rm bar, WMAP} = 0.176 \pm 0.015$ from the best-fit results
of the WMAP 3-yrs data (\cite{sp06}), a depletion parameter
$Y=0.920\pm0.023$ and the gas and stellar mass fractions
shown on the right which refer to a sample of 58 clusters
with $T_{\rm gas}>4$ keV, redshift in the range $0.14-1.26$ and discussed
in Ettori et al. (2006, in prep.). The dashed lines indicates the median value
of $f_{\rm star}=0.016$ and $f_{\rm star}/f_{\rm gas}=0.129$.}
\label{fig:1}       
\end{figure}

The X-ray emitting hot plasma in galaxy clusters represents
the 80 per cent of the total amount of cluster baryons.
It is composed by a primordial component polluted 
from the outputs of the star formation activity taking place in galaxies, 
that are the cold phase accounting for about 10 per cent
of the cluster barionic budget (see Fig.~\ref{fig:1}).

As reference, a number density of $9.77 \times 10^{-2}$
ions of helium and $4.68 \times 10^{-5}$ ions of iron is expected 
for each atom of hydrogen in a hot plasma with solar abundance
(as in \cite{ag89}; for the most recent
estimates in \cite{gs98} and \cite{a05}, the number density
relative to H is $8.51 \times 10^{-2}$ for He, 
$3.16 \times 10^{-5}$ and $2.82 \times 10^{-5}$ for Fe, respectively; see Table~1).
These values imply a mass fraction of $(0.707, 0.738)$ for H, 
$(0.274, 0.250)$ for He and $(0.019, 0.012)$ for heavier elements, 
accordingly to (\cite{ag89}, \cite{a05}), respectively.

I discuss here some speculations, and relevant implications, of the sedimentation
of helium in cluster cores (Sect.~\ref{sec:2}; details are presented in \cite{ef06})
and a history of the metal accumulation in the ICM, with new
calculations (with respect to the original work in \cite{e05})
following the recent evidence of a bi-modal distribution of the delay time in 
SNe Ia (see Sect.~\ref{sec:3}).

\section{Helium in X-ray galaxy clusters}
\label{sec:2}

Diffusion of helium and other metals can occur in the central
regions of the intracluster plasma under the attractive action 
of the gravitational potential, enhancing their 
abundances on time scales comparable to the cluster age.
For a Boltzmann distribution of particles labeled $1$ with
density $n_1$ and thermal velocity $v_{\rm th} = 
(2 kT / A_1 m_{\rm p})^{1/2}$ in a plasma with temperature $kT$ in 
hydrostatic equilibrium with a NFW potential $g(r)$, the drift velocity 
of the heavier ions $2$ with respect to $1$ is given by
(\cite{s56})
\begin{equation}
v_{\rm sed}(r) = \frac{3 m_{\rm p}^2 \ A_1 A_2 \ v_{\rm th}^3 \ g(r)}
{16 \pi^{1/2} e^4 \ Z_1^2 Z_2^2 \ n_1 \ \ln \Lambda},
\end{equation}
where $\log \Lambda$ is the Coulomb logarithm.
In \cite{ef06}, we have studied the effects of the sedimentation 
of helium nuclei on the X-ray properties of galaxy clusters.

\begin{figure}[hb]
 \vspace*{-0.3cm}
\centering
\includegraphics[height=3.7cm]{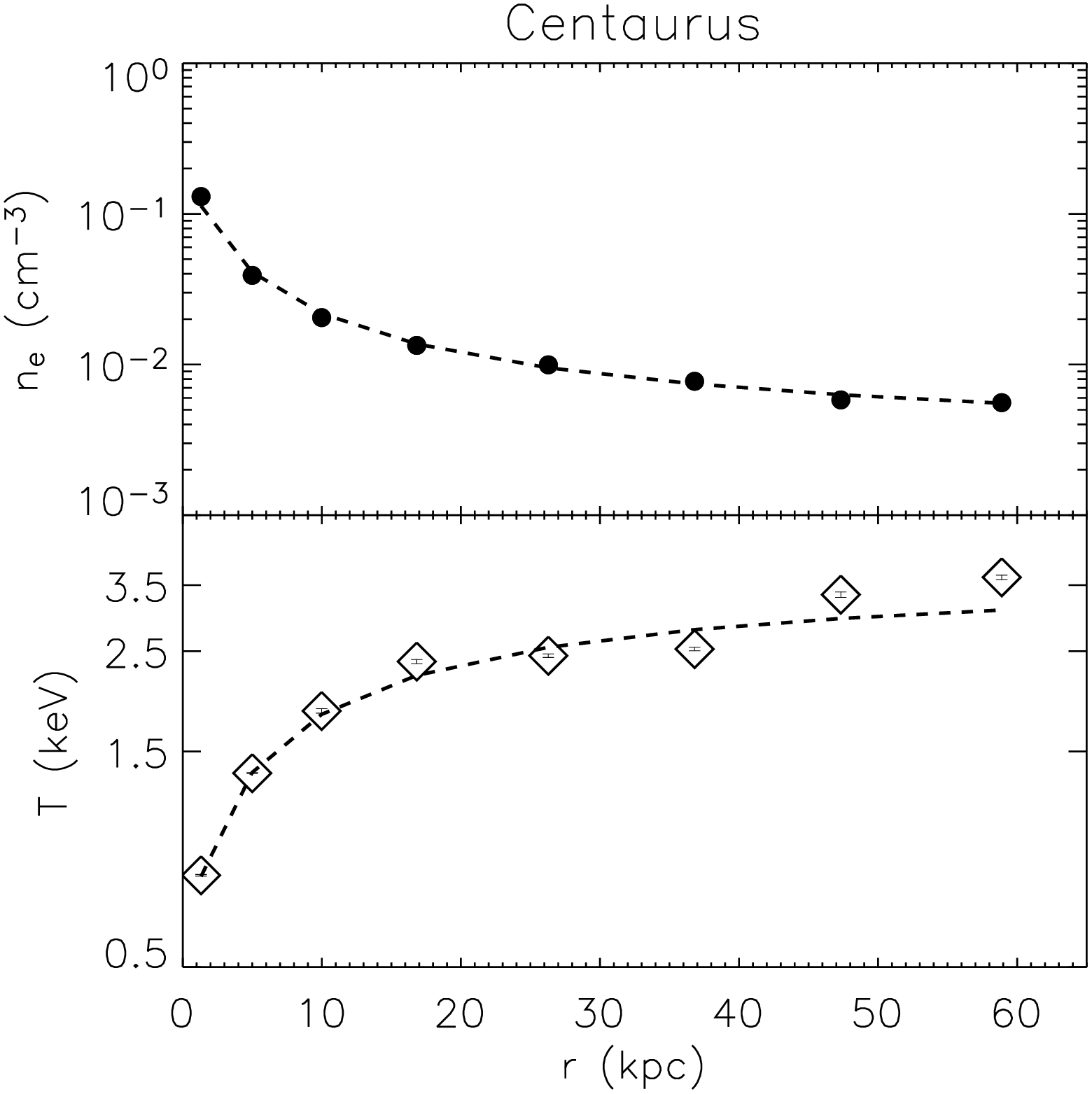}
\includegraphics[height=3.7cm]{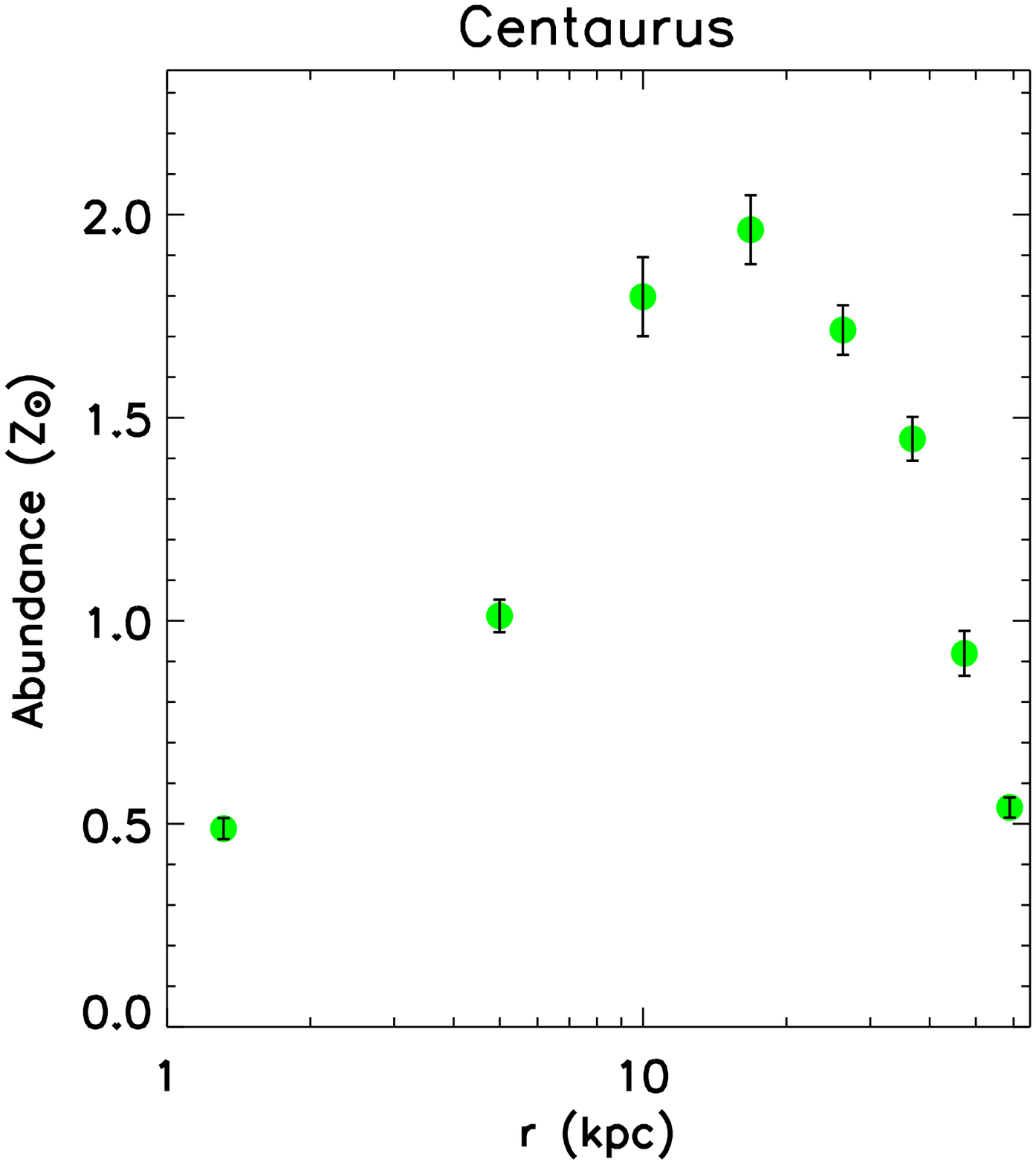}
\includegraphics[height=3.7cm]{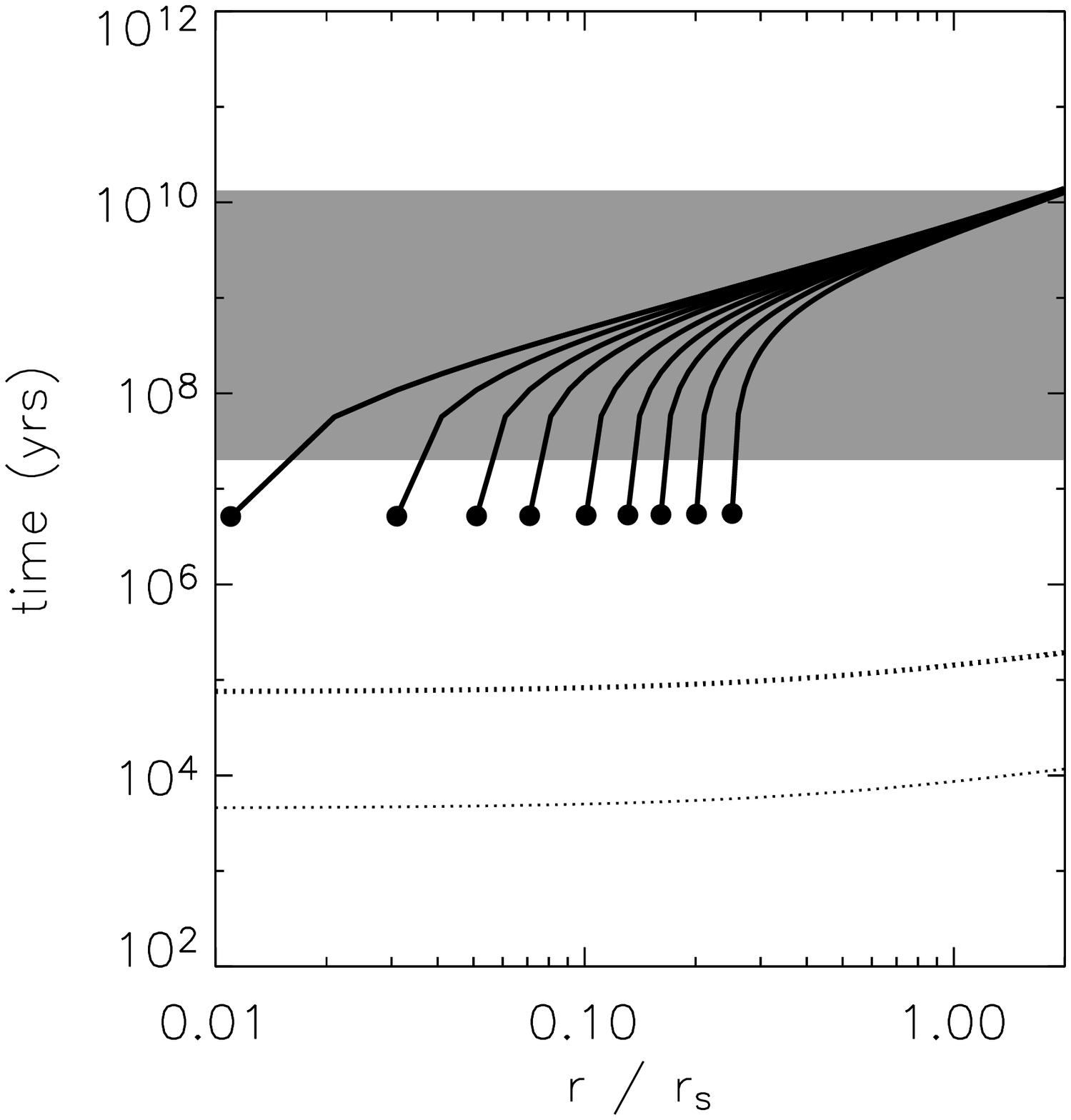}
\caption{
(Left) Best joint-fit of the gas density and temperature profile
of the deprojected data of Centaurus \index{A3526} 
(from \cite{s02}) with a modified NFW gas profile.
(Middle) Metal abundance profile with the characteristic central drop observed
in spatially well resolved cool-core clusters.
(Right) H-He (thick dotted line) and H-Fe (thin dotted line) equipartition time 
in the Centaurus cluster.
The dots mark the radius $r_{\rm in}$ at which He is accumulated from
the regions beyond with a sedimentation time represented by the solid line:
$t_{\rm sed} = \int^{r_{\rm out}}_{r_{\rm in}} dr / v_{\rm sed}(r)$.
The shaded region ranges between a $t_{\rm cool}=2 \times 10^7$ yrs and
the age of the Universe (details in \cite{ef06}).
} \label{fig:2}       
\end{figure}

We have estimated the gravitational acceleration by
fitting to the observed deprojected gas density and 
temperature profiles some functional forms that well reproduce the central 
steepening and are obtained under the assumption that the plasma
is in the hydrostatic equilibrium with a NFW potential 
(left panel in Fig.~\ref{fig:2}).
The observed gas density and temperature values do not allow
metals to settle down in cluster cores over timescales
shorter than few $10^9$ years.
On the other hand, by assuming that in the same potential
the gas that is now describing a cool core was initially 
isothermal, the sedimentation times are reduced by 
1--2 order of magnitude within $0.2 r_{200}$.
The sedimentation of helium can then take place in cluster
cores.
Even modest enhancement in the helium abundance affects 
(i) the relative number of electron and ions 
$\frac{n_{\rm e}}{n_{\rm p}} = c_M = \sum M_i Z_i N_i =
 1 +2 M_{\rm He} N_{\rm He} +\sum_{i \neq {\rm H, He}} M_i Z_i N_i$, 
(ii) the atomic mean molecular weight $\mu = 
\left(  \sum_i \frac{X_i (1+Z_i)}{A_i} \right)^{-1}$ and all the 
X-ray quantities that depend on these values, such as
the emissivity $\epsilon$, the gas mass density
$\rho_{\rm g} \propto \mu (1+c_M)$, the total
gravitating mass, $M_{\rm tot} \propto \mu^{-1}$,
the gas mass fraction, $f_{\rm gas} \propto \mu^2 (1+c_M)$.
Moreover, we show in \cite{ef06} that if we model a super-solar abundance 
of helium with the solar value, we underestimate the metal (iron) 
abundance and overestimate the model normalization or emission measure.
For example, with $M_{\rm He} = 3$, the measured iron abundance
and emission measure are $\sim0.65$ and $1.5$ times the input values,
respectively. Thus, an underestimated excess of the He abundance might explain
the drops in metallicity observed in the inner regions
of cool-core clusters (see, e.g., middle panel in Fig.~\ref{fig:2}).
It is worth noticing that the diffusion of helium can be suppressed,
however, by the action of confinement due to reasonable magnetic fields
(see \cite{cl04}), or limited to the very central region ($r<20$ kpc)
by the turbulent motion of the plasma.

\section{Iron in X-ray galaxy clusters}
\label{sec:3}
The iron abundance is nowadays routinely determined in nearby systems thanks mostly 
to the prominence of the K-shell iron line emission at rest-frame
energies of 6.6-7.0 keV (Fe XXV and Fe XXVI).  Furthermore, 
the few X-ray galaxy clusters known at $z>1$ have shown well detected Fe line, given
sufficiently long ($> 200$ ksec) {\it Chandra} and {\it XMM-Newton} exposures
(\cite{r04}, \cite{h04}, \cite{to03}).
It is more difficult to assess the abundance of other prominent metals that 
should appear in an X-ray spectrum at energies (observer rest frame)
between $\sim 0.5$ and $10$ keV, such as oxygen (O VIII) at (cluster rest frame) 
0.65 keV, silicon (Si XIV) at 2.0 keV, sulfur (S XVI) at 2.6 keV, 
and nickel (Ni) at 7.8 keV.  
In \cite{e05}, we infer from observed and modeled 
SN rates the total and relative amount of metals that should be present in the 
ICM both locally and at high redshifts.
Through our phenomenological approach, we adopt the models of SN rates as 
a function of redshift that reproduce well both the very recent observational 
determinations of SN rates at $z > 0.3$ (\cite{d04}, \cite{c05})
and the measurements of the star formation rate derived from UV-luminosity
densities and IR data sets. We then compare the products of the enrichment 
process to the constraints obtained through X-ray observations of galaxy clusters 
up to $z \sim 1.2$.

\begin{table}
\vspace*{-0.3cm}
\caption{Adopted values for the average atomic weight ($W$), solar abundance by number
with respect to H ($A$, from \cite{ag89} and \cite{a05}; $R=A_{\rm A05}/A_{\rm AG89}$), 
total synthesized isotopic mass per SN event ($m_{Ia}$ from deflagration model W7 and 
$m_{CC}$ integrated over the mass range $10-50 M_{\odot}$ with a Salpeter IMF;
see \cite{n97}) and corresponding abundance ratios by numbers with respect 
to Fe, $Y_i = m_i/(W_i A_i) \times (W_{\rm Fe} A_{\rm Fe})/m_{\rm Fe}$.
}
\centering
\begin{tabular}{c c c c c c c c c c}
\hline\noalign{\smallskip} \\
{\it metal} & $W$ & $A_{\rm AG89}$ & $A_{\rm A05}$ & $R$ & & $m_{Ia}$ & $Y_{Ia}$ & $m_{CC}$ & $Y_{CC}$\\
 & & & & & & $M_{\odot}$ & & $M_{\odot}$ \\
\noalign{\smallskip}\hline\noalign{\smallskip}
He &  4.002 & 9.77e-2 & 8.51e-2 & 0.87 & &  $-$  &  $-$  &  $-$  &  $-$  \\
Fe & 55.845 & 4.68e-5 & 2.82e-5 & 0.60 & & 0.743 &  $-$  & 0.091 &  $-$  \\
O  & 15.999 & 8.51e-4 & 4.57e-4 & 0.54 & & 0.143 & 0.037 & 1.805 & 3.818  \\
Si & 28.086 & 3.55e-5 & 3.24e-5 & 0.91 & & 0.153 & 0.538 & 0.122 & 3.526  \\
S  & 32.065 & 1.62e-5 & 1.38e-5 & 0.85 & & 0.086 & 0.585 & 0.041 & 2.284  \\
Ni & 58.693 & 1.78e-6 & 1.70e-6 & 0.95 & & 0.141 & 4.758 & 0.006 & 1.647  \\
\noalign{\smallskip}\hline
\end{tabular}
\end{table}

The observed iron mass is obtained from \cite{dg04} and \cite{to03}
(see also \cite{e04}) as
\begin{equation}
M_{\rm Fe, obs} = 4 \pi A_{\rm Fe} W_{\rm Fe} \int_0^R Z_{\rm Fe}(r) \rho_{\rm H}(r) r^2 dr,
\end{equation}
where $Z_{\rm Fe}(r)$ is the radial iron abundance relative to the solar value $A_{\rm Fe}$,
$W_{\rm Fe}$ is the iron atomic weight and 
$\rho_{\rm H}(r)$ is the hydrogen mass density.
To recover the history of the metals accumulation in the ICM,
we use the models of the cosmological rates 
(in unit of SN number per comoving volume and rest-frame year)
of Type Ia, $r_{\rm Ia}$, and core-collapse supernovae, $r_{\rm CC}$, 
as presented in \cite{d04} and \cite{s04}.
We estimate then the iron mass through the equation
\begin{equation}
M_{\rm Fe, SN} = M_{\rm Fe, Ia} +M_{\rm Fe, CC} = \sum_{dt, dV} 
\left(m_{\rm Fe, Ia} r_{\rm Ia}(dt) +m_{\rm Fe, CC} r_{\rm CC}(dt)
\right)  dt  dV 
\end{equation}
where $m_{\rm Fe, Ia}$ and $m_{\rm Fe, CC}$ are quoted in Table~1,
$dt$ is the cosmic time elapsed in a given redshift range, 
$dV$ is the cluster volume defined as the volume corresponding
to the spherical region that encompasses the cosmic background density,
$\rho_{\rm b} = 3 H_0^2 / (8 \pi G) \times \Omega_{\rm m}$ 
($\approx 4 \times 10^{10} M_{\odot}$ Mpc$^{-3}$ for the assumed cosmology),
with a cumulative mass of $M_{\rm vir} \approx 6.8 
(kT / 5{\rm keV})^{3/2} \times 10^{14} M_{\odot}$ (\cite{arn05}). 
We obtain that
these SN rates provide on average a total amount of iron that is 
marginally consistent with the value measured in galaxy clusters in the redshift 
range $0-1$, and a relative evolution with redshift that is in agreement 
with the observational constraints up to $z \approx 1.2$.  
We predict metals-to-iron ratios well in agreement with
the X-ray measurements obtained in nearby clusters
implying that (1) about half of the iron mass and
$>$ 75 per cent of the nickel mass observed locally are produced by SN Ia ejecta, 
(2) the SN Ia contribution to the metal budget decreases steeply with redshift 
and by $z \approx 1$ is already less than half of the local amount
and (3) a transition in the abundance ratios relative to the iron is present
at $z$ between $\sim 0.5$ and $1.4$, with SN CC products becoming 
dominant at higher redshifts.

\begin{figure}[ht]
\centering
\vspace*{-0.3cm}
\hbox{
\includegraphics[height=4cm]{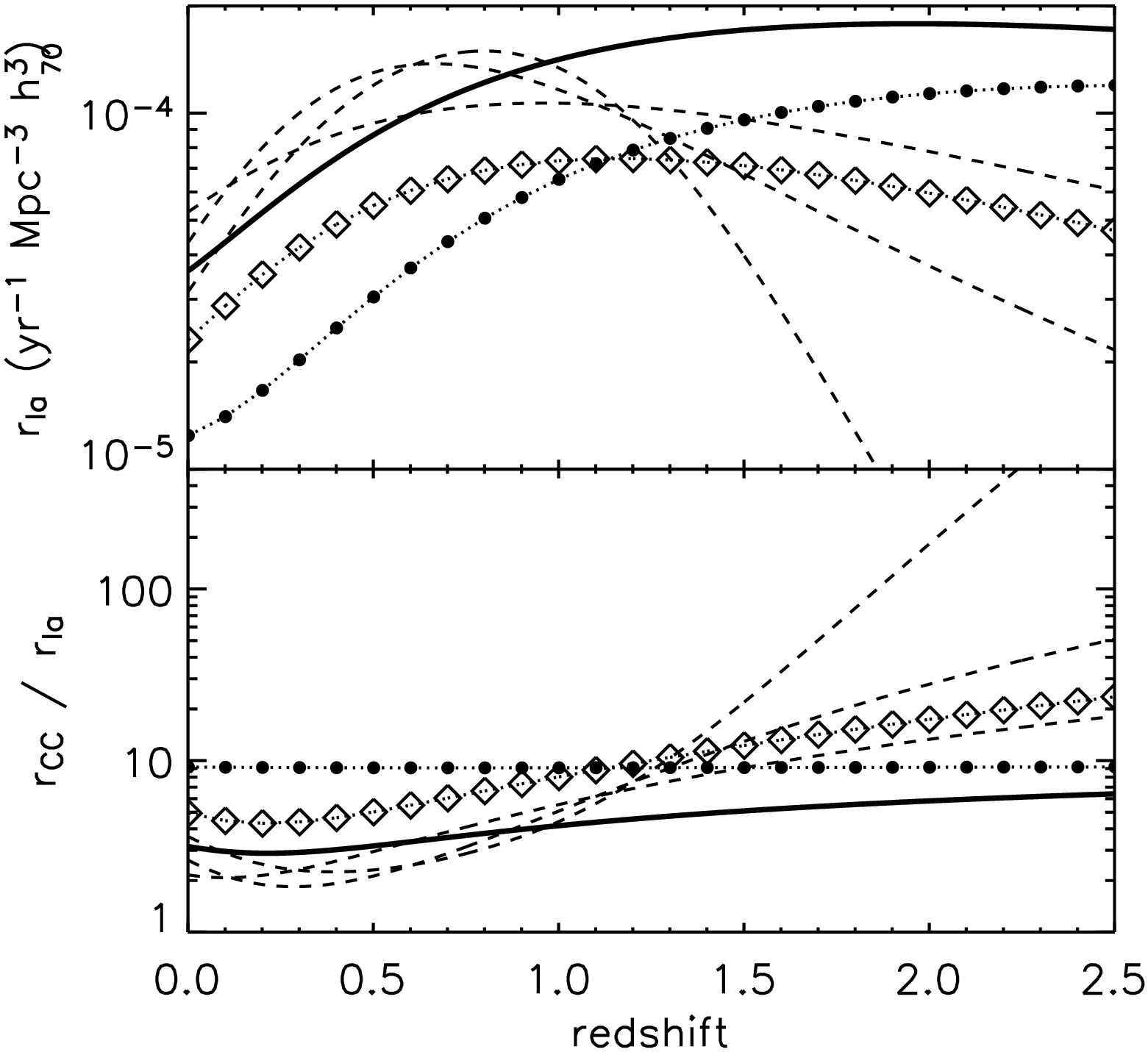}
\includegraphics[height=4cm]{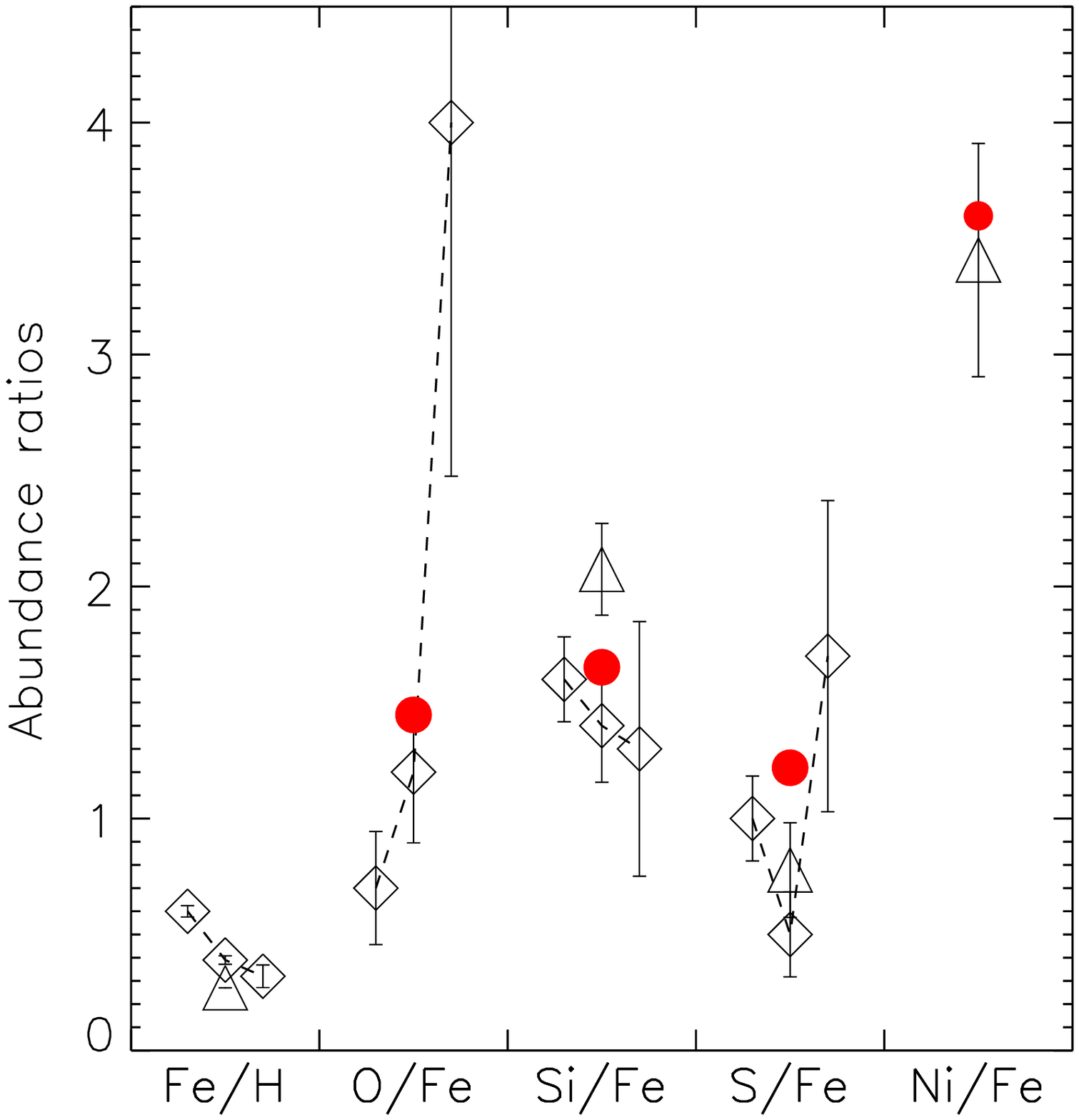}
\includegraphics[height=4cm]{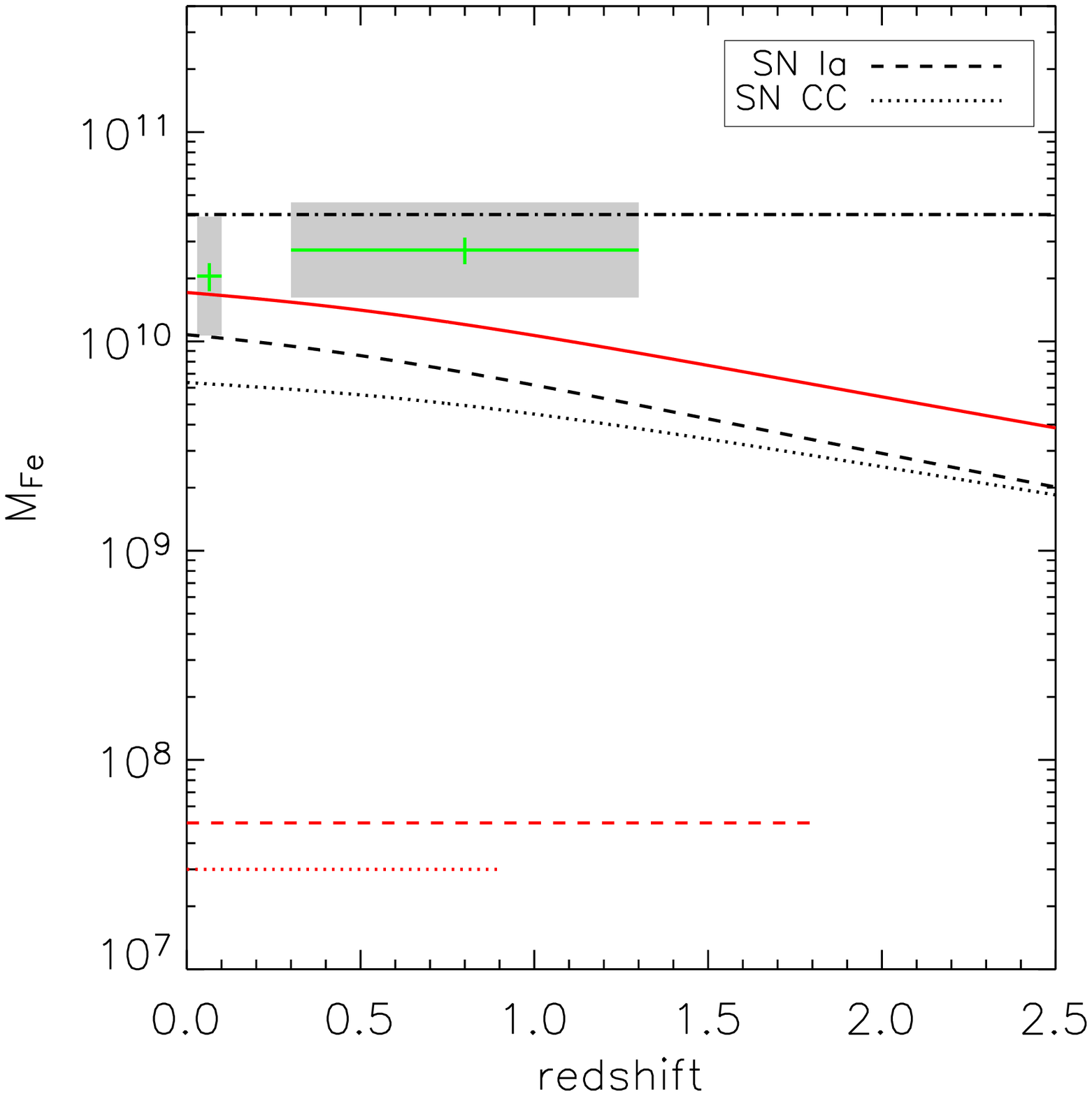}
}
\hbox{
\includegraphics[height=4cm]{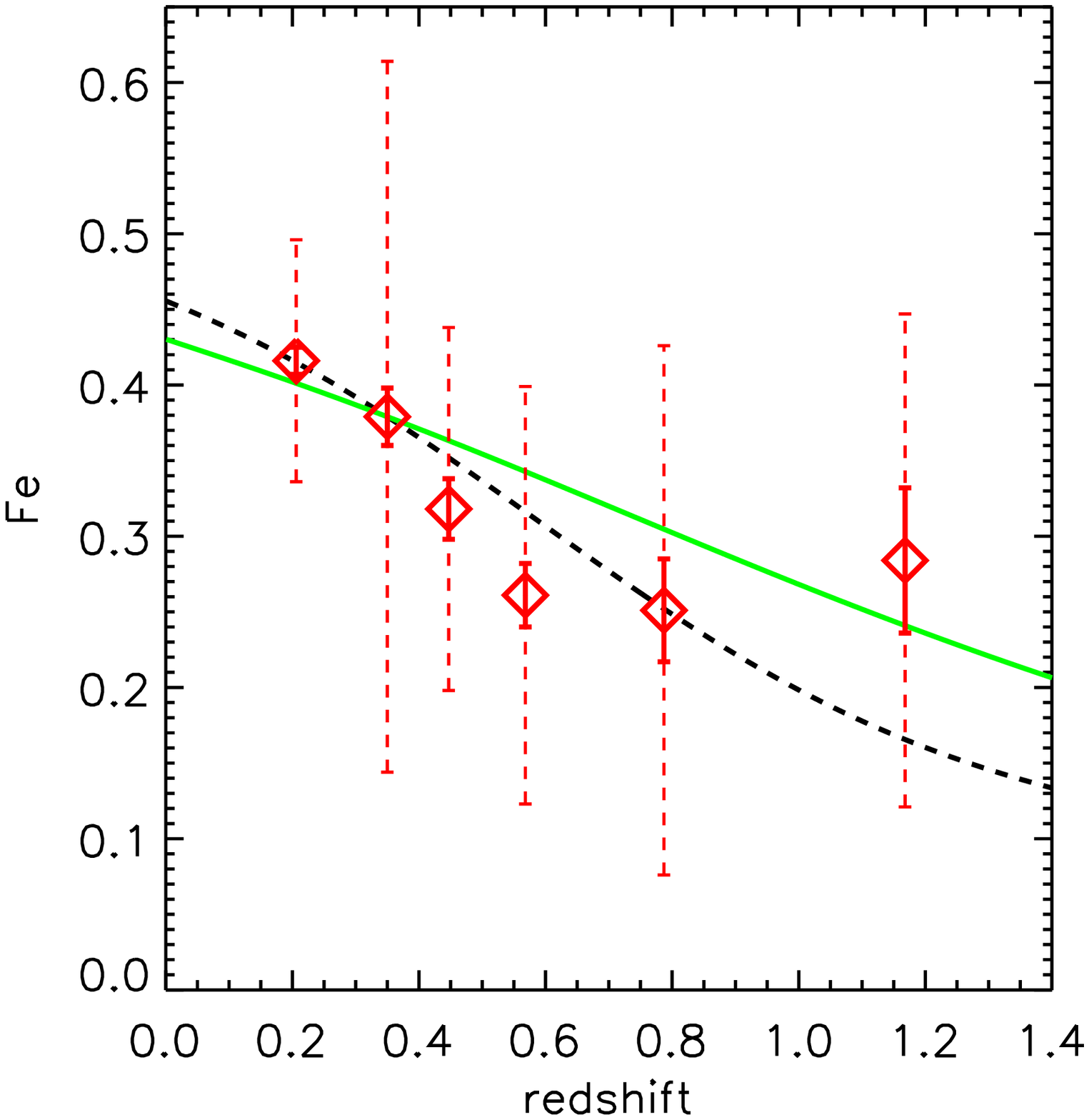}
\includegraphics[height=4cm]{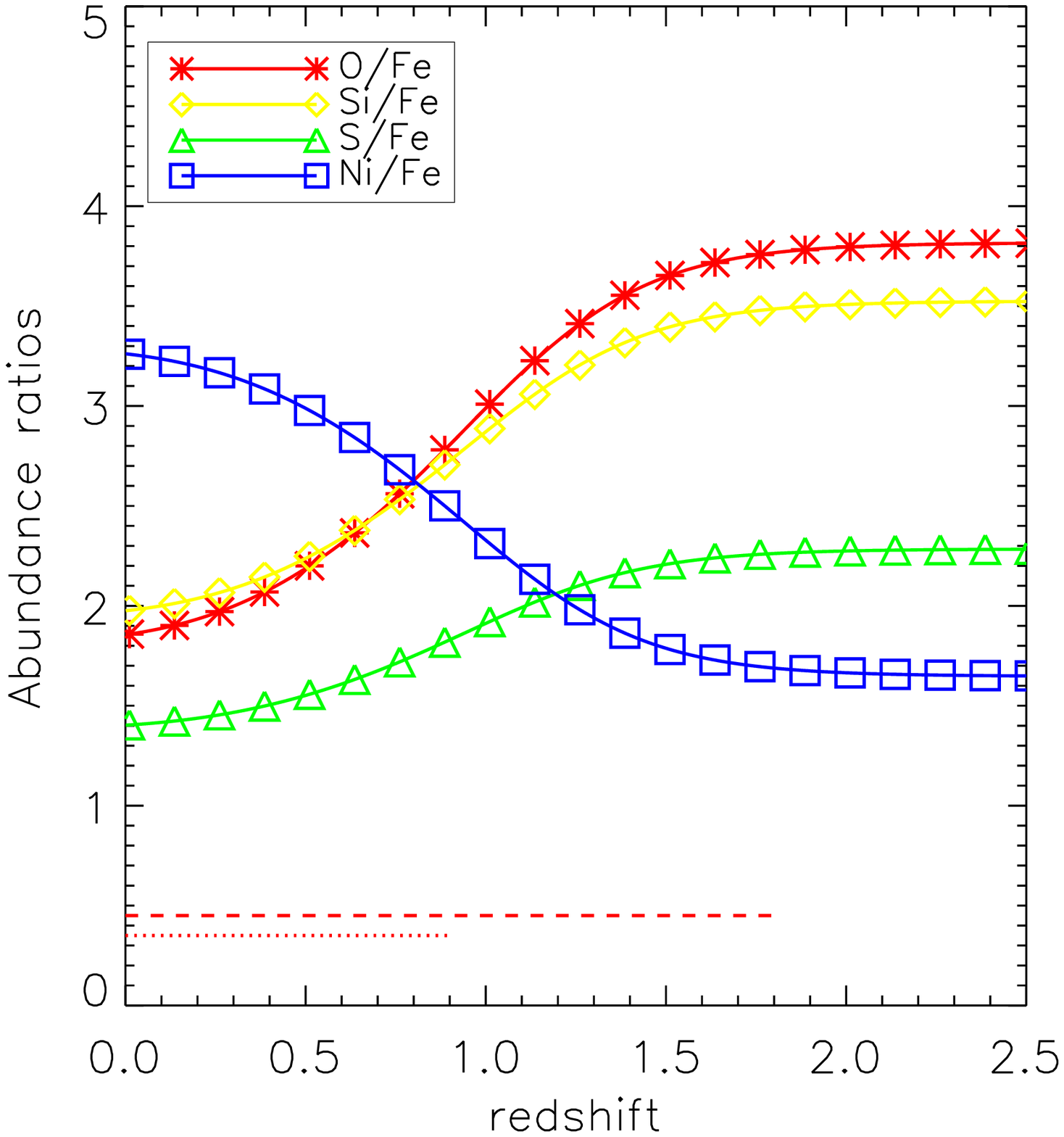}
\includegraphics[height=4cm]{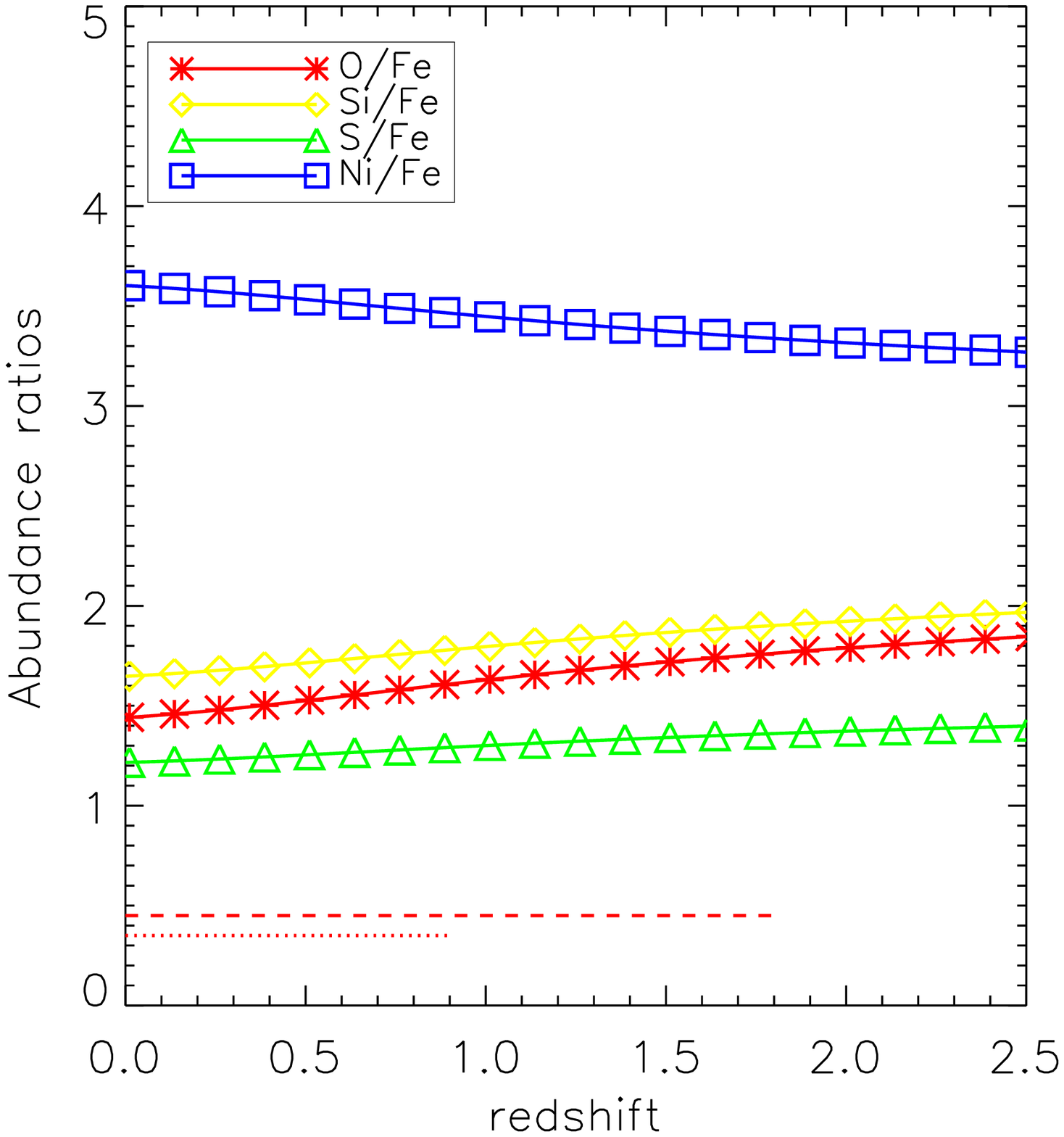}
}
%
%
\caption{
(From left to right, from up to bottom):
(A) SN Ia rates and number ratio of SNe CC to Ia as a function
of $z$ for the delay time distribution functions $\phi(t_d)$ considered in 
\cite{d04} (dashed lines) and \cite{m06} (the solid line is the sum
of the contribution from ``tardy''/squares and ``prompt''/dots population).
(B) Predictions ({\it big dots}; using $\phi(t_d)$ from \cite{m06}) 
of the local metal ratios
compared with the X-ray measurements in \cite{ta04} and \cite{b05}.
(C) Accumulation history of the iron mass, $M_{\rm Fe}$, 
as obtained from eq.~(3) with rates from \cite{m06}. 
The solid line shows the sum of SN Ia
({\it dashed line}) and SN CC ({\it dotted line}) contribution.
The accumulation history is here compared to the total iron mass measured
in samples of local (\cite{dg04}) and high$-z$ (\cite{to03})
clusters for a typical object at 5 keV.
(D) Observed (\cite{ba06}) and predicted (as in panel A and normalized to the observed
value at $z\approx0.4$) iron abundance as function of redshift.
(E-F) The last two plots show the metal abundance ratios as function of redshift
as predicted from a $\phi(t_d)$ described
by a ``narrow" Gaussian with $\tau = 4$ Gyr and $\sigma_{t_d} = 0.2 \tau$ 
as in \cite{d04} (E; see the corresponding rates in panel A) 
and from a bi-modal SN Ia distribution as in \cite{m06} (F).
Horizontal lines indicate the redshift region where the SN rates
are actually measured.
These calculations refer to the solar abundance values in \cite{ag89}
for a direct comparison with the data.
By adopting the compilation in \cite{a05}, the trends are confirmed
with the most relevant change being the predicted Ni/Fe$\approx2$ at $z=0$.
} \label{fig:3}       
\end{figure}

Recently, Mannucci et al. (2006; see also \cite{sr05}) 
argue that the data on (i) the evolution of SN Ia rate
with redshift, (ii) the enhancement of the SN Ia rate in radio-loud early type galaxies and (iii)
the dependence of the SN Ia rate on the colours of the parent galaxies suggest the existence of
two populations of progenitors for SN Ia, one half (dubbed ``prompt'') 
that is expected to occur soon after the stellar birth on time scale of $10^8$ 
years and is able to pollute significantly their environment with iron even at high redshift, 
the other half (``tardy'') that has a delay time function 
consistent with an exponential function with a decay time of $\sim3$ Gyrs. 
Due to the larger production of SN Ia at higher redshift (see panel A in Fig.~\ref{fig:3}),
the rates suggested by Mannucci et al. predict a more flat
distribution of the Fe abundance as function of $z$ than the rates tabulated in 
\cite{d04}, but still show a negative evolution partially consistent with the
measurements up to $z\approx1.2$ (panel D in Fig.~\ref{fig:3}).
Furthermore, these rates provide a better agreement than the ones in 
\cite{d04} both with local abundance ratios (panel B in Fig.~\ref{fig:3}) 
and with the overall production of Fe (panel C), that at $z=0$
has been released from SNe Ia by more than 60 per cent.
A potential discriminant is the behaviour of the abundance ratios relative to Fe in the ICM
as a function of redshift, providing the rates in \cite{m06} values of
O/Fe and Ni/Fe that are a factor of 2 different from the predictions of the
single-population delay time distribution function (see panels E and F).

%
%
%
%

%
%



\printindex
\end{document}